\documentclass[12pt]{spieman}  
\usepackage{amsmath,amsfonts,amssymb}
\usepackage{graphicx}
\usepackage{setspace}
\usepackage{tocloft}

\title{Characterization of the Teledyne COSMOS Camera: A Large Format CMOS Image Sensor for Astronomy}
\author[a,b,*]{Christopher Layden}
\author[a]{Jill Juneau}
\author[a]{Gustav Pettersson}
\author[a]{Nathan Lourie}
\author[a,c]{Benjamin Schneider}
\author[a]{Beverly LaMarr}
\author[a]{F. Elio Angile}
\author[b]{Fadi Farag}
\author[b]{Michelle Luo}
\author[b]{Zhi Zheng Ong}
\author[a]{Gabor Furesz}
\affil[a]{MIT Kavli Institute for Astrophysics and Space Research, Massachusetts Institute of Technology, 77 Massachusetts Ave, Cambridge, MA 02139, USA}
\affil[b]{MIT Department of Physics, 77 Massachusetts Ave., Cambridge, MA 02139, USA}
\affil[c]{Aix Marseille Univ, CNRS, CNES, LAM, Marseille, France}

\cftpagenumbersoff{figure} 
\cftpagenumbersoff{table} 
\begin{document} 
\maketitle

\begin{abstract}
The Teledyne COSMOS-66 is a next-generation CMOS camera designed for astronomical imaging, featuring a large-format sensor ($8120 \times 8120$ pixels, each $10\,\mu m$ wide), high quantum efficiency, high frame rates and a correlated multi-sampling mode that achieves low read noise. We performed a suite of bench-top and on-sky tests to characterize this sensor and analyze its suitability for use in astronomical instruments. This paper presents key findings, including measurements of linearity, conversion gain, read noise, dark current, quantum efficiency, image lag, and crosstalk. We found that the sensor exhibits nonlinear response at low signal levels (below 5\% of saturation). This nonlinearity is plausibly attributable to the trapping of electrons in each pixel during charge transfer. We developed and implemented a pixel-by-pixel nonlinearity correction, enabling accurate photometric measurements across the sensor's dynamic range. After implementing this correction, operating in the correlated multi-sampling mode, the sensor achieved an effective read noise of $2.9\,e^-$ and dark current of $0.12\,e^-/pix/s$ at $-25^\circ C$. The quantum efficiency exceeded 50\% for wavelengths from $250\,nm$ to $800\,nm$, peaking at 89\% at $600\,nm$. We observed significant optical crosstalk between the pixels, likely caused by photoelectron diffusion. To demonstrate the sensor's astronomical performance, we mounted it on the WINTER $1\,m$ telescope at Palomar Observatory for on-sky observations. These tests confirmed that the linearity calibration enables accurate stellar photometry and validated our measured noise levels. Overall, the COSMOS-66 delivers similar noise performance to large-format CCDs, but with higher frame rates and relaxed cooling requirements. If pixel design improvements are made to mitigate the nonlinearity and crosstalk, then the camera may combine the advantages of low-noise CMOS image sensors with the integration simplicity of large-format CCDs, broadening its utility to a host of astronomical science cases.
\end{abstract}

\keywords{CMOS image sensors; detector characterization; large-format image sensors; CCD}

{\noindent \footnotesize\textbf{*}Christopher Layden,  \linkable{clayden7@mit.edu} }

\begin{spacing}{2}   

\section{Introduction}
\label{sect:intro}  
Deep, large-scale astronomical surveys demand cameras with high resolution, large pixel scales, and low noise. These needs have historically been met with large-format charge-coupled device (CCD) image sensors with large pixels. With decades of technological maturity, a broad selection of sensitive CCDs are available with uniform, linear response that astronomers are familiar with. Over the past decade, however, Complementary Metal-Oxide-Semiconductor (CMOS) image sensors have emerged as an increasingly attractive alternative to CCDs for astronomical surveys.

CMOS image sensors inherently offer several advantages over CCDs, including higher readout rates, improved radiation tolerance, and reduced power consumption \cite{janesick_iii, janesick_iv, liu:2023}. Additionally, readout techniques such as digital multisampling enable CMOS sensors to achieve low read noise without requiring complex external readout architectures \cite{janesick_ii}. The widespread adoption of CMOS technology in consumer electronics has driven reduced costs and improved fabrication quality, allowing modern CMOS image sensors to exhibit lower dark current than typical CCDs at a given temperature. Therefore, by using CMOS image sensors, one can often avoid the cost and complexity associated with cryogenic cooling. Furthermore, with the prevalence of backside-illuminated (BSI) designs, scientific CMOS image sensors now rival or exceed CCDs in quantum efficiency within the visible spectrum. However, CMOS image sensors have yet to employ the deep depletion designs necessary to achieve competitive sensitivity to near-infrared light and X-rays.

Despite these advancements, CMOS image sensors still have downsides, such as higher nonuniformity (also referred to as fixed pattern noise, or FPN) and nonlinearity compared to CCDs \cite{janesick_ii}. This nonlinearity is typically most pronounced near saturation, where the large amount of accumulated charge alters the capacitance of the sense node, introducing significant distortion \cite{bohndiek:2008, stefanov:2022}. Additionally, most scientific CMOS image sensors have lower resolution than large-format CCDs. Therefore, when developing a survey instrument using CMOS image sensors, it is often necessary to construct a mosaic of many sensors, which can introduce gaps in sky coverage and present substantial engineering and data reduction challenges. CMOS image sensors also typically have smaller pixels than CCDs, necessitating costly re-imaging optics to critically sample the point spread functions (PSFs) observed through atmospheric seeing.

The COSMOS-66 camera (hereafter COSMOS) from Teledyne Digital Imaging (hereafter Teledyne) \cite{cheriyan:2022,teledyne_cosmos} is a promising new scientific CMOS image sensor for astronomy, combining CMOS technology with the geometry of large-format CCDs. COSMOS has an active area of approximately $8 \times 8\,cm^2$ divided into $8120 \times 8120$ pixels each of size $10\times 10\,\mu m$. COSMOS' large sensor size enables the imaging of large sky fields without the need to mosaic multiple smaller sensors, and its large pixels eliminate the need for re-imaging optics at $1\,m$-class telescopes.

\begin{table}[t!]
    \centering
    \begin{tabular}{c|c|c|c|c|c|c|c}
        & & Pixel &  & Read & Dark & Frame & Peak\\
        Name & Type & Size & Format & Noise & Current & Rate & QE \\
        & & ($\mu m$) & & ($e^-$) & ($e^-/pix/ks$) & ($fps$) & \\
        \hline
         CCD290-99\cite{chen:2022} & CCD & 10 & $9216{\times} 9232 $ & 4.5 & 0.6 ($173\,K$) & 0.08 & 92\%$^*$ \\
         CCD250-82\cite{kotov:2016} & CCD & 10 & $4096{\times} 4004 $ & 4.9 & \textless 20 ($178\,K$) & 0.5 & 96\% \\
         SIFS Skipper \cite{Villalpando_2024b}&  CCD & 15 & $6000 {\times} 1000$ & 0.5 & 0.2 ($143\,K$) & 0.005$^\dag$ & 99\% \\
         CCD201-20\cite{Harding_2016} & EMCCD & 13 & $1024{\times} 1024$ & $<1$ & $0.5$ ($188\,K$)& 26 & 95\% \\
         HWK4123 \cite{Khandelwal_2024} & CMOS & 4.6 & $4096{\times} 2304$ & 0.22 & 17 ($253\,K$) & 5$^\ddagger$ & 82\% \\
         Sony IMX455 \cite{Alarcon_2023} & CMOS & 3.76 & $9600{\times} 6422$ & 2.5 & 11 ($263\,K$) & 2.5 & 80\% \\
         GSense 400BSI \cite{Khandelwal_2024} & CMOS & 11 & $1608{\times} 1608$ & 1.8 & 400 ($248\,K$) & 60$^*$ & 95\% \\\hline
         COSMOS (spec.)\ \cite{teledyne_cosmos}& CMOS& 10 & $8120{\times} 8120$ & 0.7 & $50$ ($248\,K$) & 0.8 & $91\%$ \\
         COSMOS (meas.)\ & CMOS& 10 & $8120{\times} 8120$ & 2.9 & $120$ ($248\,K$) & 0.8 & $89\%$
         \vspace{0.01\linewidth}
    \end{tabular}
    \caption{Measured performance of various state-of-the-art CCD and CMOS image sensors that all have square pixels. Data are for each sensor's lowest read noise mode. For most of these sensors, higher frame rates can be achieved at the expense of greater read noise. Teledyne's specification and our measured results for COSMOS are included for reference. $^*$Value from manufacturer datasheet. $^\dag$Read noise of $0.5\,e^-$ is achieved over 5\% of the sensor area; read noise of $4\,e^-$ is achieved over the rest of the sensor. $^\ddagger$Frame rates up to $25\,fps$ may now be achieved with an updated version of this camera\cite{gallagher:2024}.}
    \label{tab:camera_specs}
\end{table}

Table~\ref{tab:camera_specs} compares the performance of various state-of-the-art CCDs and CMOS image sensors that have been recently evaluated for astronomy, as well as the specification values and our measured values for COSMOS. Large-format CCDs, like the e2v CCD290 used for the Wide Field Survey Telescope (WFST)\cite{chen:2022}, can achieve a read noise around $5\,e^-$ but have relatively slow readout rates. To perform a faster survey of the night sky, the the Large Synoptic Survey Telescope (LSST) uses the e2v CCD250 sensors\cite{kotov:2016} that employ 16 parallel readout registers to achieve readout rates of up to $0.5\,fps$. The SOAR Integral Field Spectrograph (SIFS) is more sensitive to readout noise and uses similarly sized Skipper CCDs, which achieve deep sub-electron read noise through repetitive non-destructive readout, though at the high cost of extremely low frame rates \cite{Villalpando_2024b,chattopadhyay:2024,Lapi_2024,tiffenberg:2017}. For instance, the SIFS Skipper CCDs require approximately 17 minutes to achieve a read noise of $0.22\,e^-$ (sufficient to resolve individual photons) over a $220 \times 1000$ pixel region. Electron-multiplying CCDs (EMCCDs), such as the CCD201-20 sensors used in the Caltech HIgh-speed Multi-color camERA (CHIMERA), can deliver high frame rates and near-negligible read noise \cite{Harding_2016,shade:2024}. However, other sources of noise (multiplication noise and spurious noise) limit the utility of EMCCDs, particularly for applications requiring a high dynamic range \cite{robbins:2003,harpsoe:2012}. All of these CCDs rely on cryogenic cooling to suppress their dark current.

Modern CMOS image sensors like the HWK4123 CMOS image sensor (commonly referred to as the ``qCMOS" and integrated into the Hamamatsu Orca-Quest C15550-20UP camera) can deliver deep sub-electron read noise, similar to a Skipper CCD or EMCCDs, but at significantly higher frame rates \cite{Khandelwal_2024,gallagher:2024}. Thanks to their fast readout, CMOS cameras can average many reads of each pixel to suppress random noise, all while maintaining similar or higher frame rates to comparable CCDs. The HWK4123 sensor has a relatively modest pixel format and size compared with typical scientific CCDs. However, sensors with more pixels or larger pixels, such as the Sony IMX455 and GSENSE 400BSI, respectively, can demonstrate read noise levels competitive with the best standard CCDs while operating at much higher readout speeds \cite{gill:2022,Alarcon_2023,Khandelwal_2024,Karpov_2020}. CMOS sensors often achieve competitive dark current at higher temperatures than CCDs and can therefore be operated with simple thermoelectric cooling systems.

Like the other CMOS image sensors in Table~\ref{tab:camera_specs}, COSMOS is specified to deliver high frame rates, low dark current without the need for deep cooling, and low read noise, while also matching the format and pixel size of the largest CCDs available on the market. In this work, we report the results of our laboratory characterization and on-sky testing of a COSMOS camera, including measurements of the nonlinearity, read noise, dark current, conversion gain, quantum efficiency, nonuniformity, image lag, and pixel crosstalk. The COSMOS sensor exhibited significant nonlinearity in the low-light regime, and we therefore applied a correction to all raw data before calculating any performance parameters. In Sec.~\ref{sect:nonlinearity_methods}, we describe the construction of this calibration and show that it delivered highly linear response across the sensor's dynamic range. In Sec.~\ref{sec:on_sky_results}, we report on the results of an on-sky demonstration with COSMOS mounted on the $1\,m$ WINTER telescope \cite{lourie:2020} at Palomar Observatory, including photometric measurements of the M38 star field. In Sec.~\ref{sect:conclusion}, we review the overall performance of the COSMOS sensor and its utility for astronomy.

\section{Methods}
\label{sect:methods}

\subsection{Camera Description and Settings}
\label{sect:setup}
The COSMOS sensor is a thinned, back-side illuminated (BSI) device and employs a five-transistor (5T) architecture with dual conversion gain, supporting multiple operational modes suitable for different astronomical purposes\cite{cheriyan:2022}. We evaluated a production demonstration unit of the COSMOS camera (serial number ENG-001), loaned to us by Teledyne. This camera had a fused silica window with a broadband-optimized anti-reflective (AR) coating, enabling sensitivity in the near-ultraviolet. We characterized the three modes shown in Table~\ref{tab:mode_summary}. We considered these modes the most relevant for astronomy: the HSHGRS and HSHGGS modes provide high frame rates, with the latter free of rolling shutter artifacts. The CMS mode achieves the lowest read noise by sampling the reset voltage level and signal voltage level eight times each, similar to the widely used correlated double sampling (CDS) technique.

\begin{table}[t]
\setlength{\abovecaptionskip}{8pt} 
    \centering
    \begin{tabular}{c|c|c|c|c|c}
         & & Mode & Bit & Maximum & Datasheet  \\
         Mode Description & Shutter & Abbreviation & Depth & Frame Rate & Read Noise* \\
         & & & & ($fps$) & ($e^-$) \\
        \hline
        Correlated Multi-Sampling & Rolling & CMS & 16 & 0.8 & 0.7 \\
        High Speed, High Gain & Rolling & HSHGRS & 14 & 18 & 1.4 \\
        High Speed, High Gain & Global & HSHGGS & 14 & 18 & 2.0 \\
    \end{tabular}
    \caption{Description of tested camera modes and the abbreviations we use for them. *See Table~\ref{tab:gain} for measured results.}
    \label{tab:mode_summary}
\end{table}

We operated the COSMOS camera at $-25^\circ C$, except when examining temperature-dependent properties such as dark current. We used a Koolance ERM-3K3UC Liquid Cooling System with 25\% concentration Propylene Glycol as coolant to remove heat from the camera's thermo-electric coolers, achieving stability within $1^\circ C$ from $-35^\circ\, C$ to $15^\circ C$. We controlled the camera using the Teledyne LightField application, automating long exposure sequences with MATLAB and Python scripts. All images were saved in FITS format. We disabled all optional correction process available in the camera firmware, including those that used optically black pixels on either side of the active sensor area to correct for row-to-row variations in bias levels as this could only be enabled together with an unknown built-in flat field correction. While the camera's firmware did not allow for saving values from the optically black pixels (using them only for immediate correction), Teledyne engineers informed us that future firmware updates are expected to enable saving these values for offline processing. 

\subsection{Descriptions of Experimental Apparatus}
\label{sect:apparatus}

\subsubsection{Uniform illumination}
\label{sect:flats_apparatus}
\begin{figure}[h]
    \centering
    \includegraphics[width=0.98\linewidth]{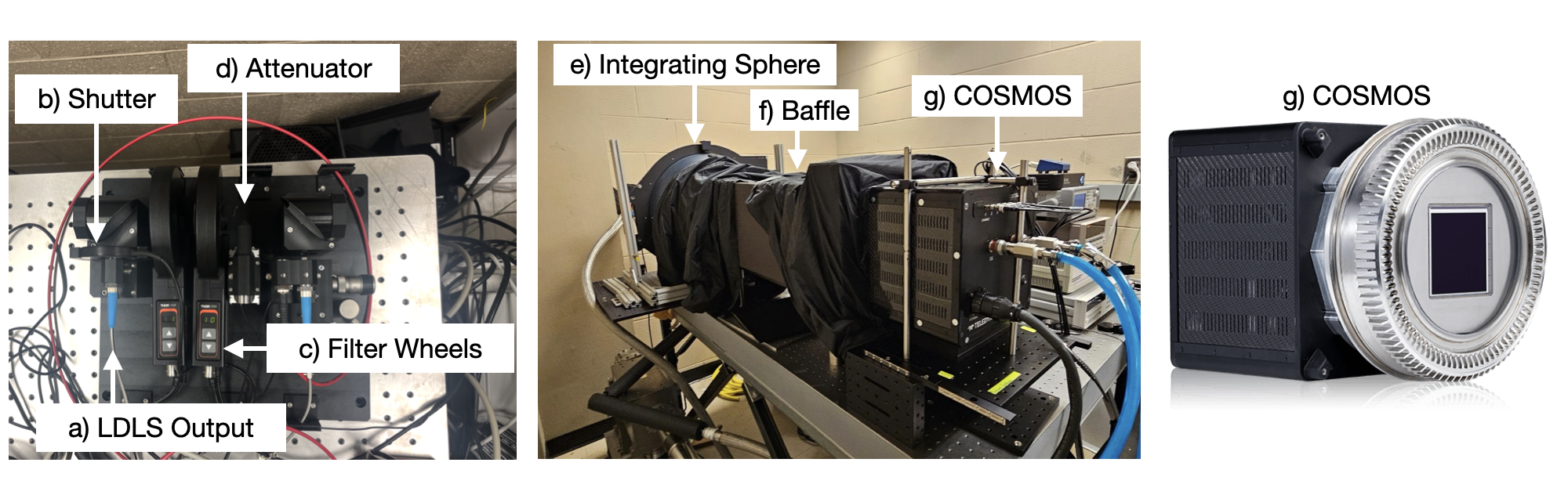}
    \caption{Optical bench setup for measuring photon transfer, nonlinearity, quantum efficiency, and uniformity. \textbf{a)} A laser-driven light source provided broadband illumination through a fiber optic cable. This cable opened to \textbf{b)}, a shutter that can be operated remotely. If the shutter was open, light passed through \textbf{c)} filter wheels to select the desired spectral band. Light could then be attenuated by \textbf{d)} a knife-edge attenuator to reach a desired illumination level. Light was then injected into \textbf{e)} an integrating sphere. \textbf{f)} A baffle of length 32 inches connected a 4 inch sphere output port to the face of \textbf{g)}, the COSMOS camera. Unseen, a calibrated photodiode was mounted on the COSMOS sensor window. The front of COSMOS is shown on the right, courtesy of Ref. \citenum{teledyne_cosmos}.}
    \label{fig:qe_setup}
\end{figure}

To study the light-sensing characteristics of COSMOS, we required a stable source of broadband illumination that is highly uniform across the entire sensor. The optical bench used for this purpose is pictured in Fig.~\ref{fig:qe_setup}. This apparatus was largely the same as that employed for evaluating CCDs for the Transiting Exoplanet Survey Satellite (TESS) \cite{Krishnamurthy_2017} and was derived from the ultra-stable light source setup developed for the CHaracterising ExOPlanets Satellite (CHEOPS) \cite{wildi:2015}. A laser-driven Xenon light source (or LDLS, namely the Energetiq EQ-99X-FC) provided stable, broadband light, and a filter wheel selected the desired passband. We used hard-coated filters from Edmund Optics, with center wavelengths ranging from $250\,nm$ to $1064\,nm$ at intervals of no more than $50\,nm$, and with full width at half maximum (FWHM) less than $10\,nm$. A knife-edge attenuator could be shifted into the beam to block a fraction of the light. The beam was then coupled into an integrating sphere with diameter $20\,in$. An output port with diameter $4\,in$ illuminated the sensor, which was placed $32\,in$ away from the port, per the EMVA Standard 1288 \cite{emva1288}. A baffle connected the port and sensor, blocking stray light. A ThorLabs FDS1010-CAL photodiode calibrated from $350\,nm$ to $1100\,nm$ by the National Institute of Standards and Technology (NIST) was mounted on the camera window. With this photodiode, we measured that the light source brightness was stable over time to within 0.2\%. When using ultraviolet light, we used a Hamamatsu S1337-1010BQ photodiode that was calibrated by the Japan Calibration Service System (JCSS) down to $200\,nm$. We precisely measured the position of the photodiode and thereby corrected for the difference in photon flux experienced at the photodiode and sensor surface. We performed these corrections and confirmed that the illumination nonuniformity over the sensor was less than 0.7\% using a simulation of the integrating sphere port geometry that agrees with results reported by the manufacturer \cite{labsphere_integrating_sphere}.

\subsubsection{Subpixel spot illumination}
\label{sect:subpixel_apparatus}
\begin{figure}[h!]
    \centering
    \includegraphics[width=0.9\linewidth]{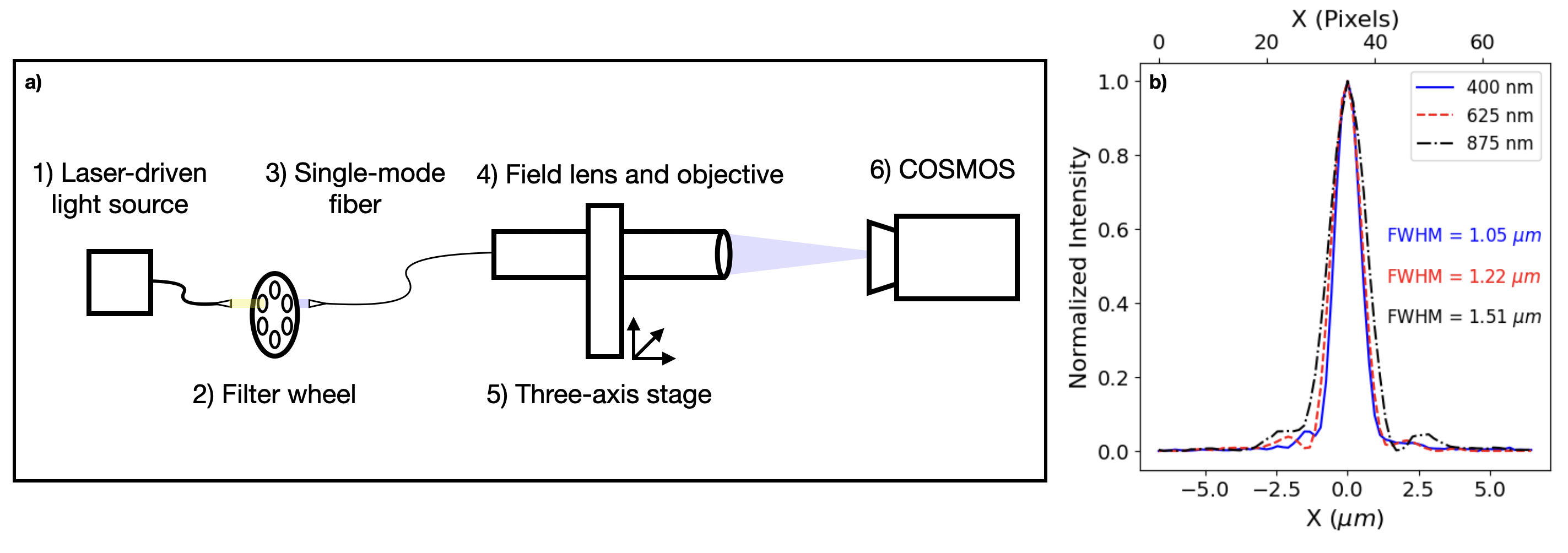}
    \caption{\textbf{a)} Diagram of the apparatus used to project a diffraction-limited spot on the COSMOS sensor. \textbf{b)} Horizontal profile of the projected spot for blue, red, and near-infrared light. Vertical profiles were nearly identical.}
    \label{fig:subpixel_setup}
\end{figure}

To study the behavior of individual pixels in COSMOS and their interactions with neighboring pixels, we projected a small spot of light that is much smaller than the $10\,\mu m \times 10\,\mu m$ pixels onto the sensor. Fig.~\ref{fig:subpixel_setup} depicts the apparatus used to generate this spot. Spectrally filtered light from an LDLS was coupled into a single-mode fiber with a mode field diameter of $\approx 5\,\mu m$. A 10x Mitutoyo Plan Apo objective and matching field lens, operating in reverse, demagnified the spot emitted by the fiber and projected it as a diffraction-limited spot on the COSMOS sensor. To verify that the spot of light was much smaller than a COSMOS pixel, we magnified and imaged the spot using a 20x Mitutoyo Plan Apo objective coupled to a FLIR Grasshopper3 GS3-U3-51S5M camera, temporarily placing these in the position of the COSMOS camera. Figure~\ref{fig:subpixel_setup}b shows the profiles of the spots for three wavelengths. All exhibited near diffraction-limited Airy disks, with FWHM below $2\,\mu m$.

\section{Image Calibration}
\label{sect:nonlinearity_methods}
\begin{figure}[h!]
    \centering
    \includegraphics[width=0.98\linewidth]{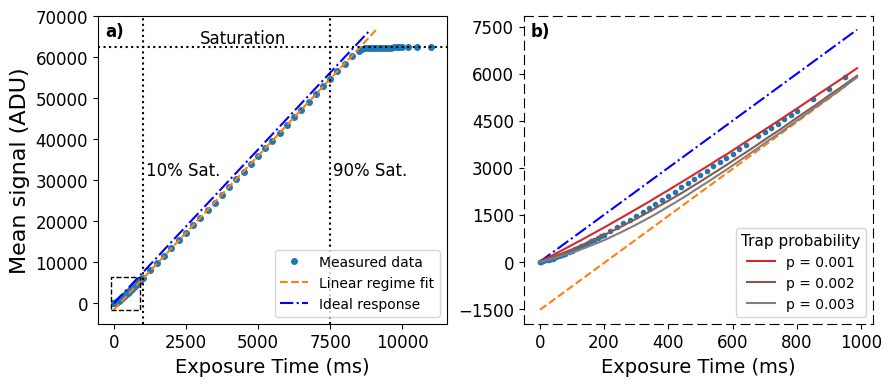}
    \caption{\textbf{a)} COSMOS response to increasing levels of illumination. Blue points show the mean signal level at each exposure time. The orange dashed line shows a fit to the sensor's linear regime (10\%--90\% saturation). The blue dashed-dotted line shows ideal sensor response, with the same slope as the orange dashed line but intersecting the origin. \textbf{b)} Zoom-in on the rectangular region marked in the left panel, showing the severe nonlinearity at low signal levels. This panel also shows the response that is expected assuming our trap model, with different values of electron capture $p$.}
    \label{fig:raw_nonlinearity}
\end{figure}

Early in our testing, we observed that COSMOS exhibited significant nonlinearity at low signal levels. It was therefore necessary to develop a calibration procedure to convert raw signal values into values that linearly increase with illumination. All evaluated modes exhibited very similar response, and the calibrations developed for them performed equally well. We constructed this calibration for each operating mode by acquiring a sequence of images at increasing exposure times using the uniform, stable light source discussed in Sec.~\ref{sect:flats_apparatus} with a filter centered at $640\,nm$. This sequence consisted of 104 exposure times, with at least 50 times chosen to yield response below 5\% of saturation. At each exposure time along this ramp, we acquired a stack of 25 frames, calculated a mean frame from this stack, and subtracted a bias frame (calculated as the mean of a stack of 50 dark exposures taken at the minimum exposure time). Fig.~\ref{fig:raw_nonlinearity} shows the mean signal value vs.\ exposure time for the CMS mode. It also shows the nonlinear response predicted by a model with a potential barrier in the photodiode, which is discussed in more detail in Sec.~\ref{sec:trap_model}. We fit the data between 10\% and 90\% of saturation to a line, shown by the orange dashed line. At low signal levels, the COSMOS response deviated severely from this line, as shown in the zoomed-in panel in Fig.~\ref{fig:raw_nonlinearity}b. Our calibration mapped the raw signal values onto the ideal linear response, formed by shifting the linear fit (orange dashed line) to pass through the origin, as represented by the blue dashed-dotted line.

\begin{figure}[h!]
    \centering
    \includegraphics[width=0.98\linewidth]{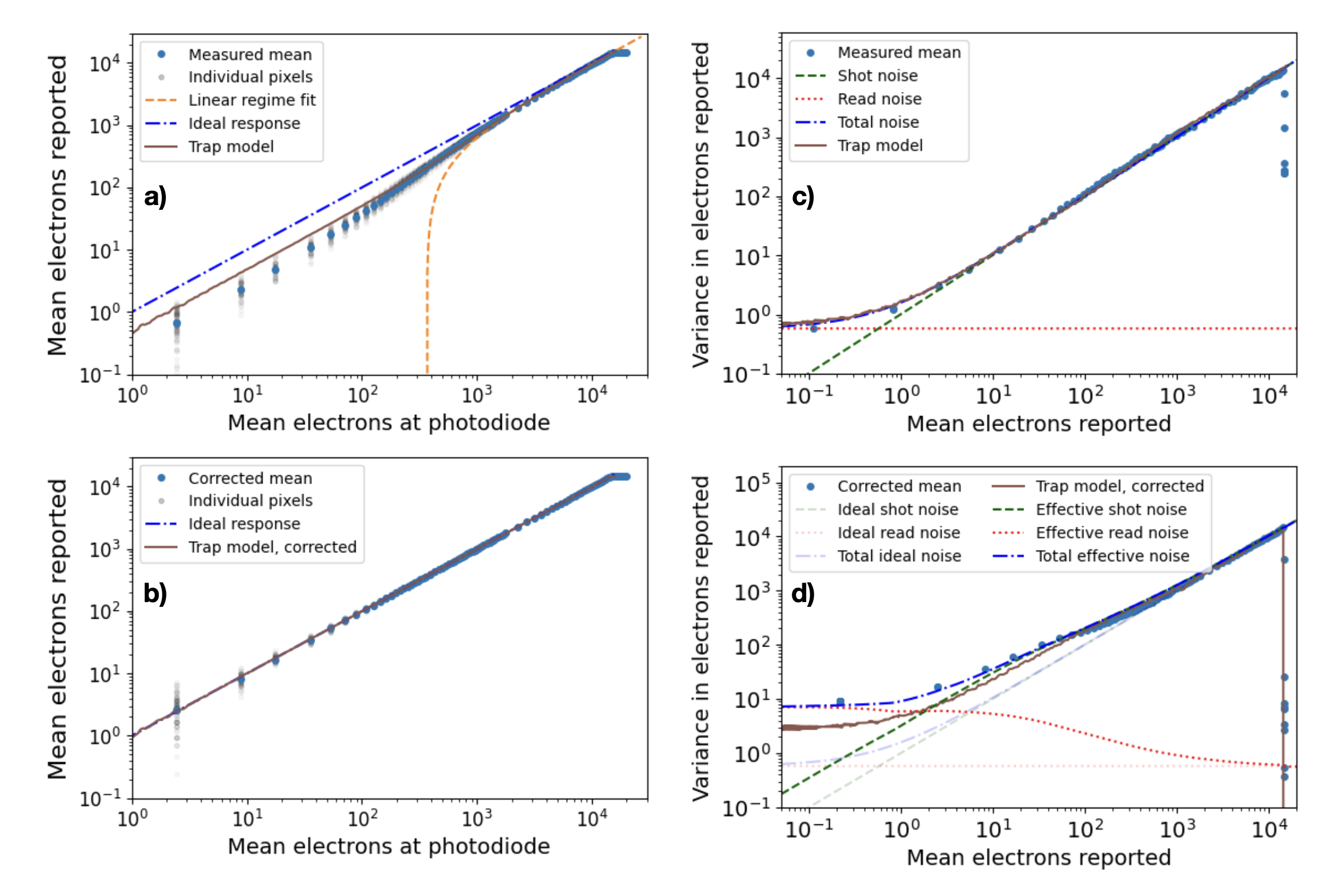}
    \caption{\textbf{a)} Raw COSMOS response to increasing illumination in CMS mode. Black markers show responses of 50 individual pixels; blue dots show their mean. The orange dashed curve shows the best fit line to the linear regime, and the blue dashed-dotted line this same fit shifted to pass through the origin. The brown solid curve shows the response predicted by the best fit trap model with $N_t=365$, $p=0.002$. \textbf{b)} Response after applying the linearity calibration to the COSMOS data and to the trap model. \textbf{c)} PTC measured for raw COSMOS data (blue markers) and predicted for the best fit trap model (brown solid curve). Variance from read noise is shown in the red dotted line, from shot noise in the green dotted line, and from the sum of these two in the blue dashed-dotted line. \textbf{d)} PTC measured for corrected COSMOS data and corrected trap model data. Variance from effective read noise is shown in the red dotted line, from effective shot noise in the green dashed line, and from the sum of the two in the blue dashed-dotted line. Semi-transparent curves show the components of noise expected from an ideal sensor with a read noise of $0.7\,e^-$.}
    \label{fig:nonlinearity}
\end{figure}

The raw nonlinear response and the ideal response to which we calibrate are also shown in log-log scale in Fig.~\ref{fig:nonlinearity}a. Here we have converted the y-axis into units of reported electrons by dividing by the sensor gain (measured via X-ray illumination as discussed in Sec.~\ref{sect:xray_results}), and we have converted the x-axis into units of electrons at the photodiode by multiplying exposure time values by the slope of the orange best fit line (which is the product of the electron flux, the quantum efficiency, and the gain) and dividing by the gain. Figure~\ref{fig:nonlinearity}a also shows in black semi-transparent points the measured response for individual pixels, demonstrating a strong spread in the response curves for different pixels. It was therefore necessary to develop a calibration for each individual pixel. We constructed this calibration using polynomial fits, as follows: we fit the measured data points from 0\% to 5\% saturation to an $11^{th}$ order polynomial. We fit the measured data points from 5\% to 100\% saturation (plus a transition region of two points extending below 5\%) to a $5^{th}$ order polynomial. In the transition region, the corrected value is taken as the weighted average of the two polynomials. These fits were computationally efficient, smooth, and monotonically increasing between measured values (with possible exception in the transition region, where a given pixel may exhibit nonlinearity smaller than $0.5\%$).

Figure~\ref{fig:nonlinearity}b shows the corrected sensor response, which fell directly along the ideal response. The spread in the responses of individual pixels was significantly reduced, with any residual spread attributable to shot noise and read noise. To confirm that the nonlinear response is stable across time, we repeated illumination ramps with 11 frames at each exposure time on subsequent days and verified that our calibration also linearized this fresh data. We optimized the application of the correction to require one second per frame, if using a reasonably powerful consumer graphics processing unit (GPU). This may not be fast enough for real-time processing of high frame rate observations. Throughout this paper, unless otherwise noted, it should be assumed that any COSMOS images have been processed by this pixel-by-pixel nonlinearity calibration.

\subsection{Modeling the Nonlinear Response}
\label{sec:trap_model}
Teledyne engineers have identified the root cause of the nonlinearity and have shared simulation results with us (under non-disclosure agreement) that matched our experimental results well. We have been informed that Teledyne is implementing a solution to overcome the low level non-linearity that will be employed for future versions of the COSMOS sensor \cite{Teledyne2025}.

Here, we discuss a model we have developed that may plausibly explain the nonlinear response. In Fig.~\ref{fig:raw_nonlinearity}, the slope of the sensor response with respect to exposure time is smaller at low signal levels, indicating that such signal levels are underestimates of the true illumination level. Such underestimation could be caused by the trapping of electrons in each pixel, preventing some electrons from reaching the sense node. We adopted a simple, one-dimensional model of this trapping to study whether this mechanism could explain our observations. In our model, each pixel has $N_t$ charge traps. When an electron encounters a trap, it can be captured with probability $p$, causing the trap to become occupied and the number of traps to decrease. Each electron collected at a pixel photodiode is passed over each unoccupied trap, and we count the number of uncaptured electrons. We calculate the electrons reported at the node as this number of uncaptured electrons plus a noise term (sampled from a Gaussian distribution with standard deviation of $0.7\,e^-$). We repeated this random process many times to find the mean and variance in the number of electrons reported at the sense node for a Poisson-distributed number of electrons with mean $N_e$ was collected at the photodiode.

In this model, when a large number of electrons $N_e \gg N_t$ are collected at the photodiode, $N_e-N_t$ electrons are reported at the sense node, corresponding to all traps being filled. If this model applies to COSMOS, we can calculate the average number of traps each pixel, $N_t$, by finding the x-intercept of the best fit line to the linear regime. We found $N_t\approx365$, as shown in Fig.~\ref{fig:nonlinearity}a. We then observed whether the COSMOS response matches what is expected from our model with $N_t=365$ and some value of $p$. We show these expected response curves in the Fig.~\ref{fig:raw_nonlinearity}b. These model response curves captured the shape of the COSMOS response curve fairly well, with a value of $p=0.002$ yielding the best agreement across the sensor's dynamic range, although as shown in Fig.~\ref{fig:nonlinearity}a it slightly overestimates small signal values ($<1\%$ of saturation) and underestimates moderate signal values ($1-10\%$ of saturation). Given this reasonable agreement and the simplicity of our model, we found it plausible that a mechanism involving charge trapping could explain the nonlinearity.

In Fig.~\ref{fig:nonlinearity}c, we show for the raw COSMOS response the variance in the number of electrons reported vs.\ the mean number of electrons reported. This curve is referred to as a photon transfer curve, or PTC. We calculated the variance across the 25 frames at each exposure time. Despite COSMOS' nonlinear response, this PTC looked normal, with the variance being exactly what one would expect from read noise at $0.7\,e^-$ (red dotted line) and shot noise (green dashed line). Figure~\ref{fig:nonlinearity}c also shows the PTC expected if the camera's response was given by our best fit trap model with $N_t=365$ and $p=0.002$. This PTC also appeared normal, except for a very small bump in variance beginning near $N_t$ (this bump was more pronounced for larger values of $p$), showing further agreement with our observations.

We applied the same calibration to the response predicted for our trap model with $p=0.002$ that we did to our measured values. As shown in Fig.~\ref{fig:nonlinearity}c, this calibration worked equally well for the trap model data. In Fig.~\ref{fig:nonlinearity}d, we show the PTC for the corrected COSMOS data and the trap model data. For both of these, after correction, the variance increased significantly up to approximately $3N_t$ reported electrons, whereafter the original variance was measured. These changes occur because our nonlinearity correction scales up low signal values and their associated noise to compensate for the electrons that fail to reach the sense node. 

Importantly, the sensor's non-linearity and the need for implementing a correction calibration results in the specified physical read noise of $0.7\,e^-$ being scaled by around a factor of $4$ to approximately $3\,e^-$, and the shot noise being scaled up by the square root of this factor for low signals. Both this \textit{effective} read noise and \textit{effective} shot noise taper to their original values around $3N_t$. All data in this work reports the effective noise due to the nonlinearity, because noise is most relevant at low light levels where the nonlinearity is present. The read noise and shot noise for the trap model data did not scale up as much as the measured data, because this model delivered response closer to linearity at very low signal levels than COSMOS.

\section{Results}
\label{sect:results}

\subsection{Photon Transfer}
\label{sect:ptc_results}

We constructed photon transfer curves (PTCs) to measure the conversion gain of each sensor mode and to confirm that the noise at a given signal level can be estimated as the sum of the effective read noise and effective shot noise discussed in Sec.~\ref{sec:trap_model}. We used the same illumination ramp procedure to make these PTCs that we used to construct the nonlinearity calibration (see Sec.~\ref{sect:nonlinearity_methods}. For each exposure time, we calibrated all of the frames, then calculated the mean signal and variance for each pixel across the stack. We then found the average mean signal and variance across the sensor. Figure~\ref{fig:nonlinearity}d shows the PTC for the CMS mode, after the mean and variance had been rescaled by the conversion gain measured via X-ray illumination. The PTCs for the other modes were very similar. Before scaling to units of electrons, we fit the linear region of the PTC (between 20\% and 70\% of saturation), where the effective shot noise from nonlinearity agrees well with ideal shot noise. The slope of this line provided the conversion gain of the sensor, in $ADU/e^-$. As discussed in Sec.~\ref{sect:crosstalk_results}, we corrected the conversion gain measurement for the small effect of interpixel capacitance. Values for the extracted conversion gains for each mode are given in Table~\ref{tab:gain}. Because the linear fits may have been slightly affected by the effective noise from nonlinearity, and because a correction for interpixel capacitance was required, we used conversion gains measured from X-ray illumination (discussed in Sec.~\ref{sect:xray_results}) for our analyses.

\subsection{X-Ray gain measurements}
\label{sect:xray_results}
\begin{figure}[h!]
    \centering
    \includegraphics[width=0.98\linewidth]{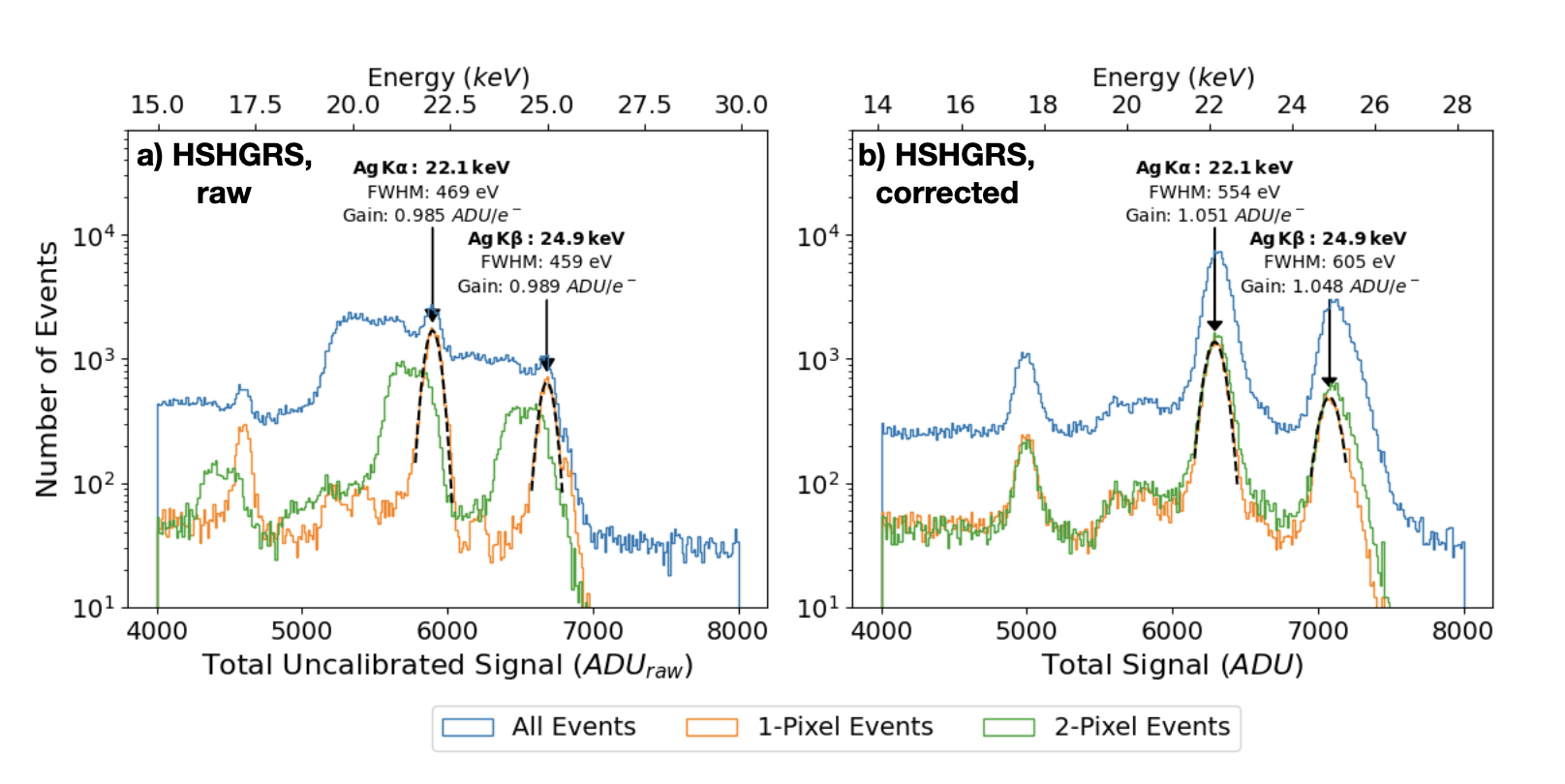}
    \caption{X-ray spectrum of a cadmium-109 source as measured by COSMOS for \textbf{a)} raw data in HSHGRS mode, and \textbf{b)} the same data with the nonlinearity correction applied. Orange curves show events for which no adjacent pixels satisfied the split threshold. Green curves show events in which two pixels satisfied the split threshold. Blue curves show all events, regardless of how charge was shared.}
    \label{fig:xray_spectra}
\end{figure}

As a second method to determine the conversion gain of the COSMOS sensor, we positioned a cadmium-109 radioactive source near the sensor's window, then shielded the camera with blackout curtains to block stray light. This source emitted X-rays with prominent lines at energies of $22.103\,keV$ and $24.935\,keV$, corresponding to $K\alpha$ and $K\beta$ doublets from silver-109, the primary decay product of cadmium-109 \cite{xraydatabooklet}. We collected frames at an exposure time of $5\,s$ for 2 hours for the HSHGRS and HSHGGS modes and for 45 minutes for the CMS mode. For each mode, we constructed X-ray spectra using both raw and calibrated data. To do so, we identified events as local maxima exceeding a $1000\,ADU$ threshold, and we calculated the pulse height for each event by summing the signals from pixels in a $5 \times 5$ region that met a split threshold defined as four times the measured read noise. Events were graded based on the number of pixels above the split threshold. We show these raw and calibrated spectra for the HSHGRS mode in Fig.~\ref{fig:xray_spectra}. The histograms for the CMS and HSHGGS modes were nearly identical, with x axes scaled by each mode's conversion gain. Due to nonlinearity errors, the peaks for multi-pixel events in the uncalibrated spectrum were wider and had lower pulse heights than for single-pixel events. The calibrated spectra showed good agreement between single-pixel and multi-pixel events, validating the nonlinearity correction.

We calculated conversion gains by fitting Gaussians to the $22.1\,keV$ and $24.9\,keV$ lines in the single-pixel event spectra, assuming an average of $E/3.69\,eV$ electrons are generated in silicon at $-25^\circ\,C$  by an X-ray of energy $E$\cite{lowe:2007}. We found the overall conversion gain as the average of the values from the two peaks. As summarized in Table~\ref{tab:gain}, conversion gains measured from X-ray illumination agreed closely with those measured from photon transfer.

\begin{table}[h!]
    \centering
    \begin{tabular}{c|c|c|c|c|c}
        & Conversion Gain &  Conversion Gain & RMS Effective & Mean Dark & Full Well \\
        Mode & (X-ray) & (PTC) &  Read Noise & Current & Capacity \\
        & ($ADU/e^-$) & ($ADU/e^-$) & ($e^-$) & ($e^- /pix/s$) & ($e^-$)\\
        \hline
        CMS & $4.30\pm 0.02$ & $4.28\pm 0.06$& $2.9\pm0.2$ & $0.12\pm 0.01$ & 14500 \\
        HSHGRS & $1.05\pm 0.01$ & $0.99\pm 0.03$ & $6.3\pm0.2$ & $0.11\pm 0.01$ & 15400 \\
        HSHGGS & $1.06\pm 0.01$ & $1.05\pm 0.03$ & $5.9\pm0.2$ & $0.11 \pm 0.01$ & 14800
        \vspace{0.01\linewidth}
    \end{tabular}
    \caption{The electron conversion gains measured via X-ray illumination and photon transfer, the RMS effective read noise, and the mean dark current at $-25^\circ C$ for the COSMOS operating modes of interest. Conversion gains measured via photon transfer have been corrected for the effects of interpixel capacitance.}
    \label{tab:gain}
\end{table}

\subsection{Defect Pixel Characterization}
\label{sect:defect_results}
The term ``defect pixel'' encompasses pixels with high dark current (``hot pixels''), low sensitivity (``lazy pixels''), or anomalous response that prevented calibration. We identified uncalibratable and lazy pixels during nonlinearity calibration, flagging pixels that exhibited non-monotonic response to increasing photon flux, that had sudden spikes in response, or that had not reached $50\%$ saturation when the rest of the sensor had reached full saturation. As shown in Table~\ref{tab:defect_pixels}, the number of such pixels was very small for the CMS and HSHGGS modes, but for the HSHGGS mode, nearly the entire top 75 rows exhibited non-monotonic response at low levels of illumination. To identify pixels with anomalous behavior that went unnoticed in calibration, we measured the conversion gains of all pixels using their photon transfer curves. We flagged pixels with extreme conversion gains using 
\(5\sigma\) clipping.

\begin{figure}[h!]
    \centering
    \includegraphics[width=0.98\linewidth]{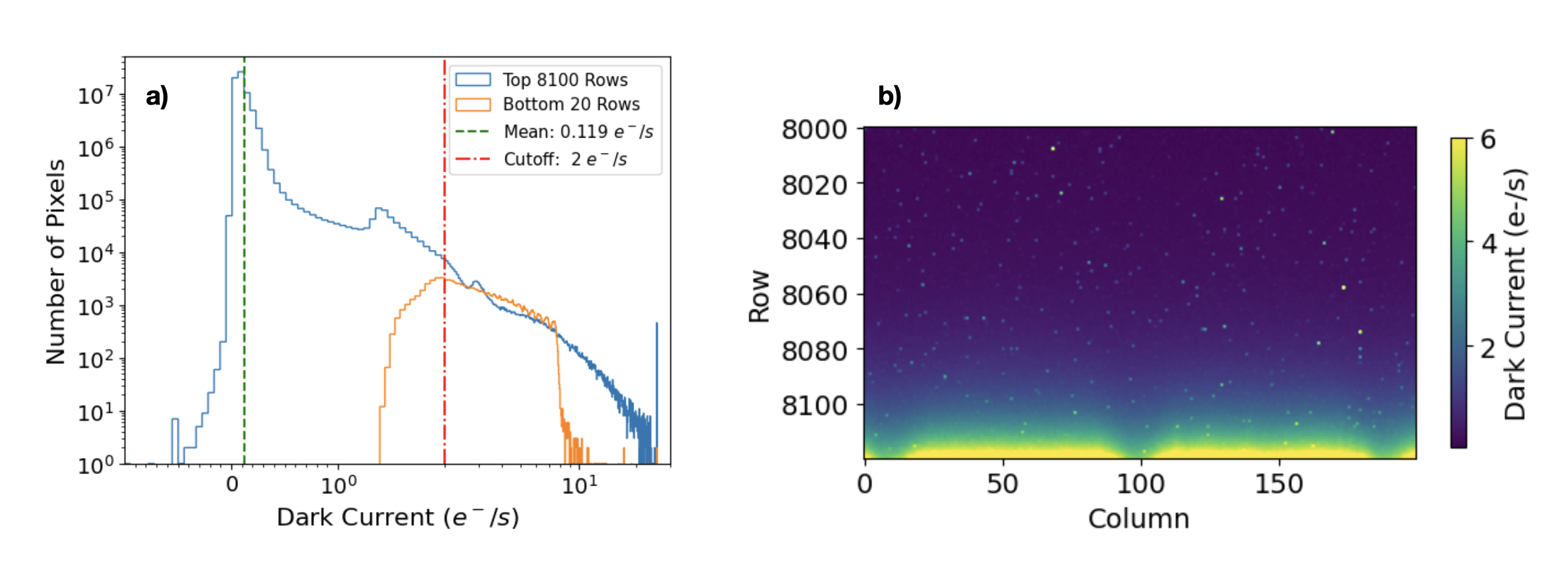}
    \caption{\textbf{a)} Dark current in individual pixels for the CMS mode. The top 8100 rows and the bottom 20 rows are plotted as separate populations. \textbf{b)} Dark current at the bottom edge of the COSMOS sensor, demonstrating a glowing pattern with a period of $\approx 90\,$ pixels.}
    \label{fig:dark_defect_hist}
\end{figure}

We identified hot pixels as those with an average dark current greater than $2\,e^-/s$, as measured in 10-minute dark exposures at $-25^\circ C$. Because more or less conservative thresholds may be desired, we show the distribution of pixel dark current values in Fig.~\ref{fig:dark_defect_hist}a. As expected, the dark current measured in each pixel was nearly identical for each mode. Most of the pixels in the bottom 20 rows of the sensor were hot pixels, likely because of amplifier glow, as shown in Fig.~\ref{fig:dark_defect_hist}b. As these rows can easily be excluded in their entirety, in Table~\ref{tab:defect_pixels} we report the fraction of defect pixels in the top $8100\times 8120$ section of the sensor. We excluded defect pixels when calculating sensor performance parameters.

\begin{table}[t]
    \centering
    \begin{tabular}{c|c|c|c|c}
        Mode & Uncalibratable/Lazy & PTC Bad Pixel & Hot Pixel & Total Defect Pixel  \\
        & Pixel Fraction (\%) & Fraction (\%) & Fraction (\%) & Fraction (\%)\\
        \hline
        CMS & 0.017 & 0.015 & 0.206 & 0.238 \\
        HSHGRS & 0.024 & 0.011 & 0.228 & 0.263 \\
        HSHGGS & 1.096 & 0.011 & 0.231 & 1.338
        \vspace{0.01\linewidth}
    \end{tabular}
    \caption{Fraction of defect pixels in each of the COSMOS operating modes of interest, for the top $8100$ rows in the sensor (out of $8120$ total). Hot pixels were defined as having mean dark current above $2\,e^-/s$ at $-25^\circ C$. Nearly all of the pixels in the bottom 20 rows were hot pixels.}
    \label{tab:defect_pixels}
\end{table}

\subsection{Read Noise}
\label{sect:read_noise_results}
\begin{figure}[h]
    \centering
    \includegraphics[width=0.98\linewidth]{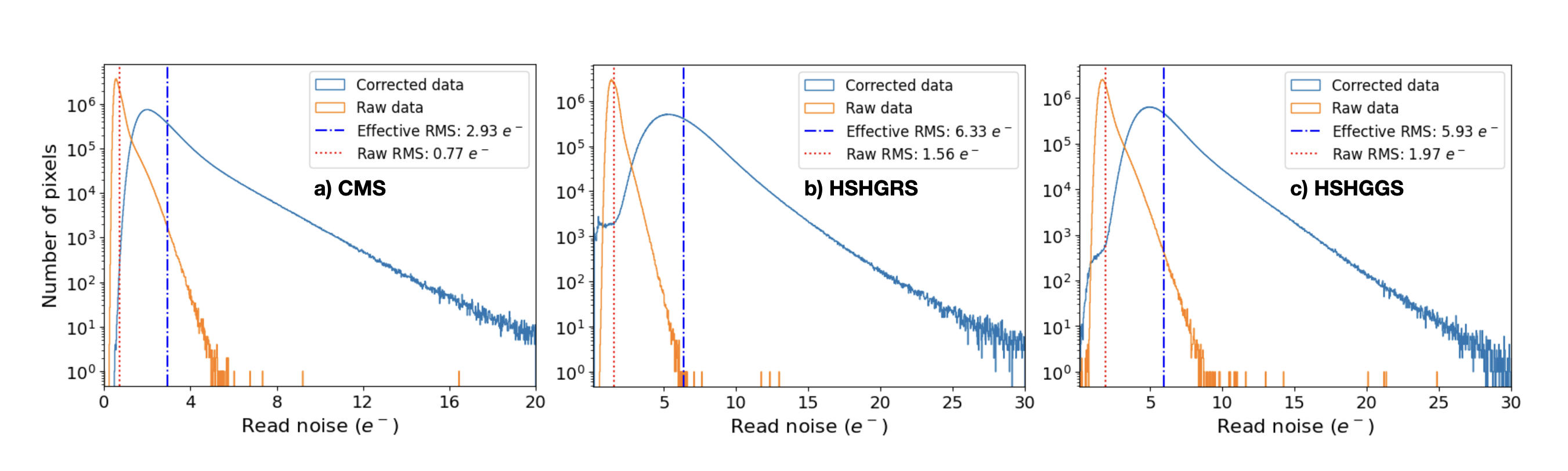}
    \caption{Histograms showing read noise in each pixel for the \textbf{a)} CMS mode, \textbf{b)} HSHGRS mode, and \textbf{c)} HSHGGS mode.}
    \label{fig:read_noise_hists}
\end{figure}

To measure the sensor's read noise, we captured a stack of 50 dark bias frames, each with the minimum exposure time for the given mode. We calculated the standard deviation of the signal level for each pixel across the stack. We computed the root mean square (RMS) of these per-pixel standard deviations and divided by the conversion gain measured via X-ray illumination. This calculation yielded the read noise, expressed in electrons ($e^-$), which we report in Table~\ref{tab:gain}. We performed this calculation using raw bias frames to calculate the physical read noise from the camera's electronics and using calibrated bias frames to calculate the effective read noise. As discussed in Sec.~\ref{sec:trap_model}, this effective read noise is larger than the physical read noise due to scaling by the nonlinearity calibration. Figure~\ref{fig:read_noise_hists} shows the distribution of per-pixel read noise measurements recorded in raw and calibrated bias frames for each mode.

\subsection{Dark Current}
\label{sect:dark_current_results}
\begin{figure}[h]
    \centering
    \includegraphics[width=0.5\linewidth]{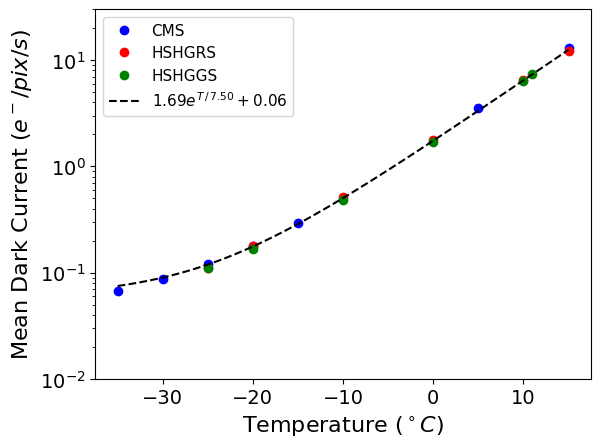}
    \caption{Mean dark current across the COSMOS sensor vs.\ temperature for the three modes of interest.}
    \label{fig:dc_vs_temp}
\end{figure}

We measured dark current by capturing dark frames at temperatures ranging from $-35^\circ C$ to $15^\circ C$ at an exposure time of $10$ minutes. We determined dark current by dividing the mean signal value by the conversion gain and the exposure time. We observed minimal variation between the dark current for the three different operating modes, as seen in Fig.~\ref{fig:dc_vs_temp} and Table~\ref{tab:gain}. We found that the dark current was well fit by an exponential function with a doubling temperature of $5.2^\circ C$ plus an asymptotic value of $0.06\,e^-/pix/s$.

\subsection{Quantum Efficiency}
\label{sect:qe_results}
\begin{figure}[h]
    \centering
    \includegraphics[width=0.6\linewidth]{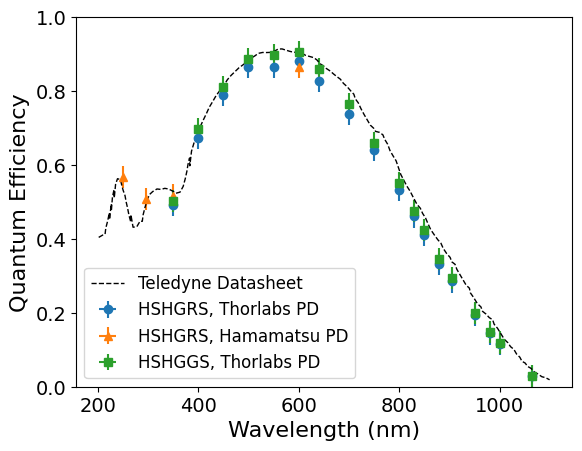}
    \caption{Measured quantum efficiency of COSMOS. Blue circles: data measured using ThorLabs photodiode and the HSHGRS mode. Orange triangles: data measured using Hamamatsu photodiode and the HSHGRS mode. Green squares: data measured using the ThorLabs photodiode and HSHGGS mode. Black dotted line: specification provided by Teledyne  \cite{teledyne_cosmos}. Error bars equal to the maximum difference between overlapping data points.}
    \label{fig:qe}
\end{figure}

We measured the quantum efficiency of COSMOS between $250\,nm$ and $1064\,nm$ using the uniform light source discussed in Sec.~\ref{sect:flats_apparatus}. For each wavelength, we selected an exposure time that achieved approximately 50\% of the sensor's saturation. At each setting, we captured three frames with the shutter open and three with the shutter closed. Simultaneously, we recorded the measurements from the calibrated photodiode to determine the incident photon flux. To calculate the quantum efficiency at each wavelength, we used the average nonlinearity-corrected and dark-subtracted signal from the sensor, the photon flux derived from the photodiode measurements, and the sensor’s conversion gain measured via X-ray illumination. The resulting QE curve is shown in Fig.~\ref{fig:qe}. The sensor's QE exceeded 50\% over the wavelength range from $250\,nm$ to $800\,nm$, peaking at $89\pm2\%$ around 600\,nm. 

\subsection{Nonuniformity}
\label{sect:nonuniformity_results}

\begin{figure}[h]
    \centering
    \includegraphics[width=0.98\linewidth]{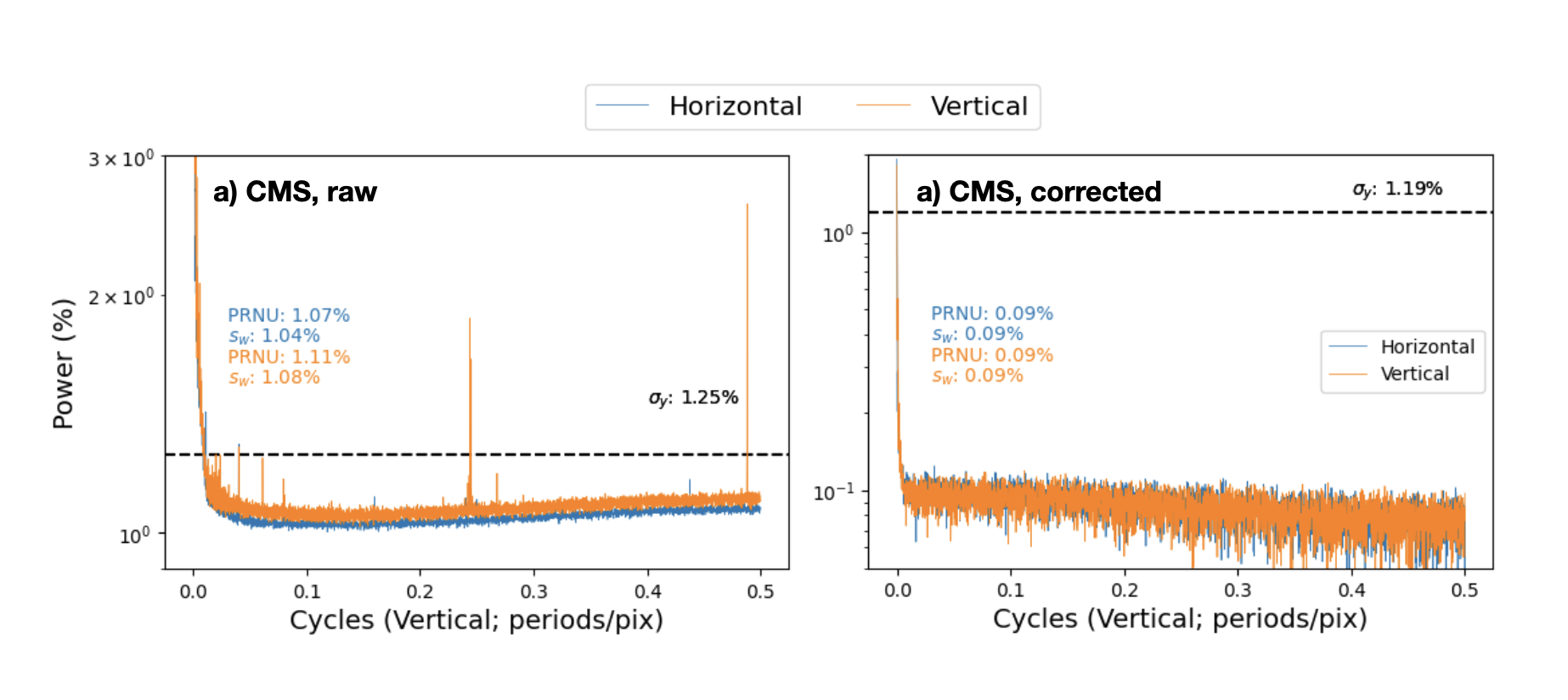}
    \caption{Horizontal and vertical spectrograms for \textbf{a)} raw images and \textbf{b)} calibrated images. The pixel response nonuniformity PRNU is calculated as the mean spectrogram value, while white noise ($s_w$) is determined from the median. The dashed black line marks the mean value of the temporal variance in the response of a pixel, $\sigma_y$, predominately from shot noise.}
    \label{fig:spectrograms}
\end{figure}

We analyzed the spatial variance in COSMOS' pixel response, commonly referred to as fixed pattern noise (FPN), by capturing a stack of 25 frames illuminated to $50\%$ saturation and 25 corresponding dark frames at the same exposure time. We calculated the pixel response nonuniformity (PRNU) for both uncalibrated and calibrated images using the equations outlined in the EMVA1288 standard. For the CMS mode, the measured PRNU was $1.1\%$ for raw images and $0.09\%$ for calibrated images. The results were consistent for the HSHGRS and HSHGGS modes. Additionally, we generated spectrograms for the average raw image and the average calibrated image, as shown in Fig.~\ref{fig:spectrograms}. These spectrograms revealed that the calibration effectively eliminated all periodic spatial variations, as evidenced by the absence of spikes in the spectrograms. Moreover, the calibration substantially reduced non-periodic variations, as reflected by the decreased level of white noise ($s_w$).

\begin{figure}[h!]
    \centering
    \includegraphics[width=0.98\linewidth]{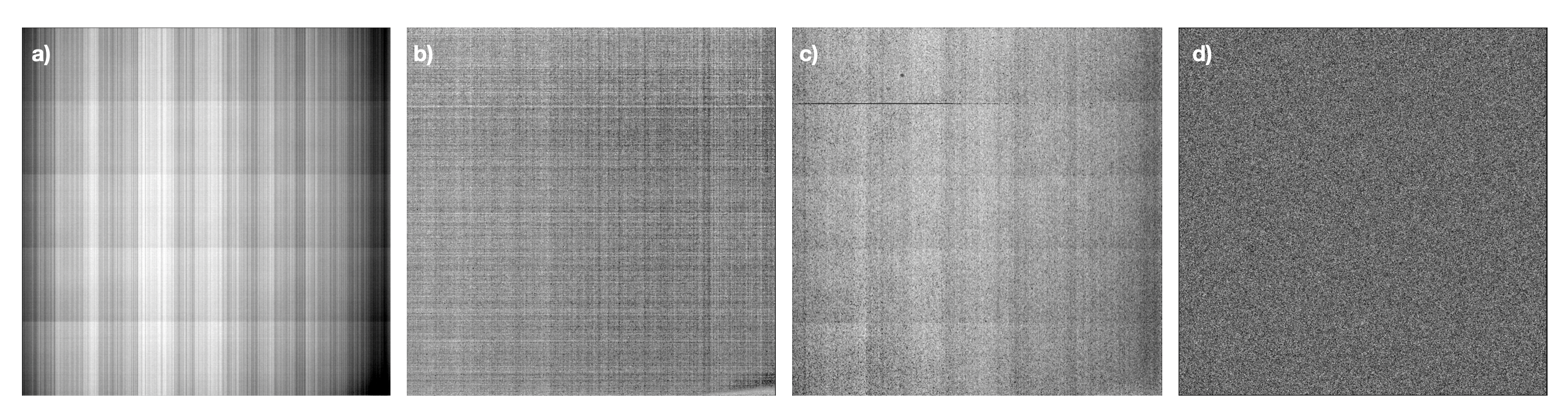}
    \caption{Fixed pattern noise before and after calibration for the HSHGRS mode. \textbf{a)} An uncalibrated bias frame. \textbf{b)} The same bias frame, calibrated. \textbf{c)} An uncalibrated flat field frame, with an average of $500\,e^-/pix$. \textbf{d)} The same flat field frame, calibrated.}
    \label{fig:cosmos-frames}
\end{figure}

Figure~\ref{fig:cosmos-frames} demonstrates visually the levels of nonuniformity before and after calibration for the HSHGRS mode. Figures~\ref{fig:cosmos-frames}a and \ref{fig:cosmos-frames}b show uncalibrated and calibrated bias frames, respectively. The calibration removed most of the structure in the bias frame, but some row-to-row and column-by-column correlations remain. These correlations could likely be removed using optically black pixels, which we could not access. Figures~\ref{fig:cosmos-frames}c and \ref{fig:cosmos-frames}d show uncalibrated and calibrated frames, respectively, when the sensor collected an average of $500\,e^-/pix$. All structure in the uncalibrated frame is removed by the calibration.

\subsection{Pixel Crosstalk}
\label{sect:crosstalk_results}
The sharing of signal between neighboring pixels, referred to as signal crosstalk, degrades the photometric performance of an image sensor by effectively broadening the point spread function (PSF) projected onto the sensor. This phenomenon is commonly parametrized in optical engineering literature as a sensor's modulation transfer function (MTF). In CMOS image sensors, crosstalk can primarily arise from two mechanisms: diffusion of photo-generated electrons and capacitive coupling, also known as interpixel capacitance (IPC).

Crosstalk from IPC correlates the noise between neighboring pixels, causing traditional mean-variance methods (such as fitting to a PTC curve) to underestimate the sensor's conversion gain. This can propagate to an overestimate of the sensor quantum efficiency~\cite{moore:2006}. IPC is more pronounced in sensors with deep depletion. Crosstalk from diffusion occurs when photoelectrons generated in one pixel spread to neighboring pixels before being collected. Diffusion crosstalk leaves the noise between pixels uncorrelated. Several techniques can mitigate diffusion crosstalk: pixels can be depleted all the way to the illuminated surface to ensure photoelectrons drift immediately; microlenses can focus light more efficiently into the depletion region; deep trench isolation can physically block diffusion between pixels; and high resistivity in the epitaxial layer can decrease electron mobility. The COSMOS sensor employs a high-resistivity epitaxial layer and an optimized back-thinning process to reduce the size of the non-depleted region\cite{Teledyne2025}, but it may still exhibit some level of diffusion crosstalk.

\begin{figure}[h!]
    \centering
    \includegraphics[width=0.98\linewidth]{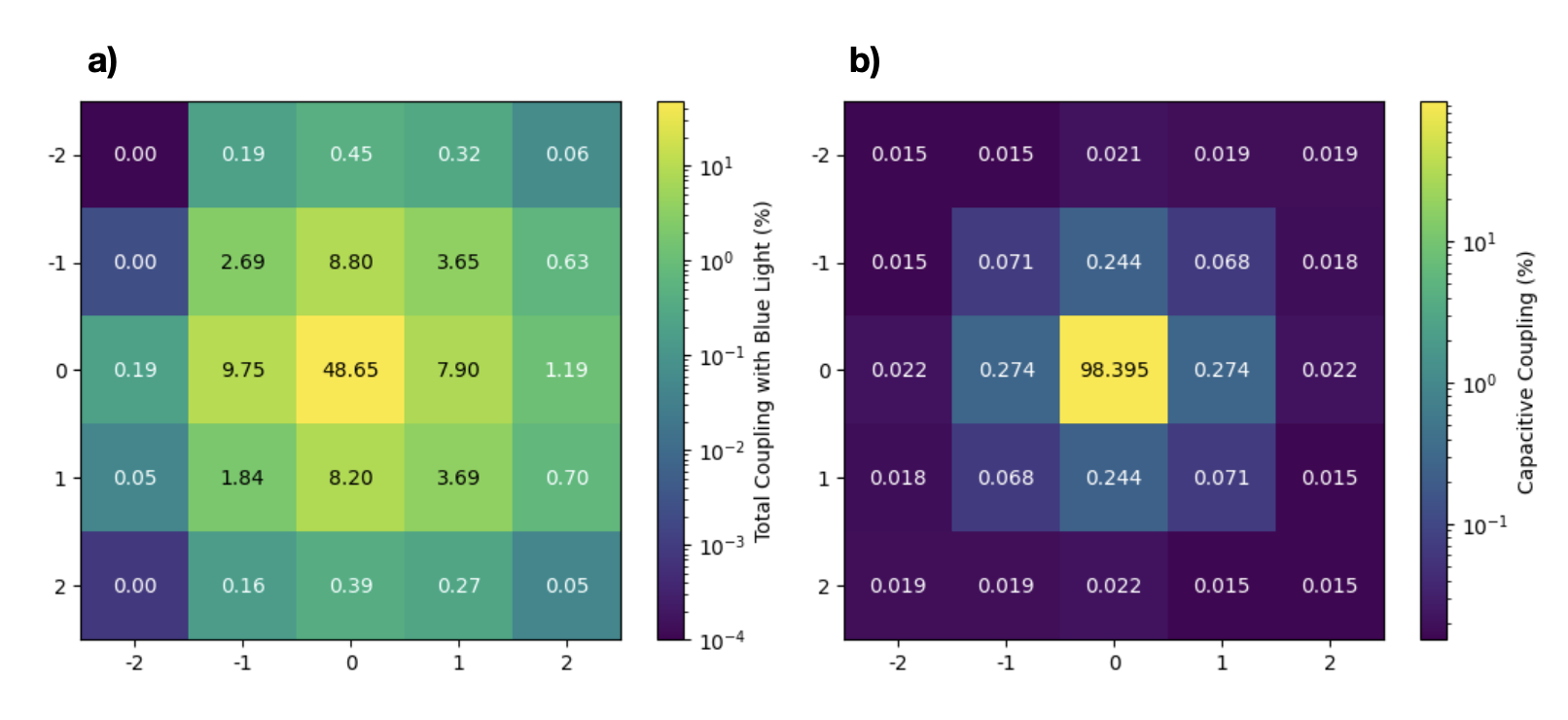}
    \caption{Coupling between COSMOS pixels. \textbf{a)} Total crosstalk between an illuminated pixel (center) and the neighboring $5\times 5$ pixel region, for blue light, the HSHGRS mode, and an exposure time of $1\,s$. The value at each pixel shows the fraction of the total signal that was recorded in that pixel when the central pixel is illuminated by the light spot. \textbf{b)} IPC kernel, showing the fraction of capacitive coupling between neighboring pixels. Coupling was nearly negligible even in adjacent pixels, so IPC was not the primary mechanism for the observed crosstalk.}
    \label{fig:coupling_heatmaps}
\end{figure}

To measure overall level of signal crosstalk in COSMOS, we illuminated an individual pixel using the subpixel spot apparatus described in Sec.~\ref{sect:subpixel_apparatus} and imaged the surrounding $11\times 11$ pixel region of interest (ROI). We ensured that the spot was focused on the sensor by maximizing the fraction of the signal reported by the central pixel relative to the total signal in this ROI. We aligned the spot in the center of a pixel by adjusting the spot's position until the four adjacent pixels reported similar signal values. We used real-time raw images to make these adjustments, but the signal values of the central pixel and its immediate neighbors were all in the linear regime. We then captured three exposures at three increasing illumination levels, for three different wavelengths ($400\,nm$, $625\,nm$, and $875\,nm$) and two different modes (CMS and HSHGRS). We used exposure times of $0.1\,s$, $0.5\,s$, and $1\,s$ for the increasing illumination levels, employing neutral density filters to ensure that these exposure times yielded approximately 6\%, 30\%, and 60\% saturation, respectively, for each mode and wavelength. For each configuration, we calculated the fraction of the signal captured by the central pixel relative to the entire ROI. We repeated this test using different pixels across the sensor and confirmed that the results were independent of pixel position. The results are shown in Table~\ref{tab:crosstalk_data} and for one configuration in Fig.~\ref{fig:coupling_heatmaps}.

We observed substantial crosstalk across all configurations: the central pixel fraction remained around or below 50\%, despite receiving more than 95\% of the incident light. Crosstalk levels were invariant across operating modes and exposure times, showing no evidence of the ``brighter-fatter effect'' seen in CCDs~\cite{Antilogus_2014}. However, crosstalk depended on wavelength, with more pronounced effects for blue and near-infrared light.  This wavelength dependence is consistent with diffusion crosstalk: blue light ($400\,nm$) is absorbed near the sensor's surface, farther from the depletion region, increasing the likelihood of charge diffusion into neighboring pixels. Near-infrared light ($875\,nm$), with an absorption length exceeding $10\,\mu m$, often passes through the sensing region and diffusely reflects off circuitry into neighboring pixels. When X-rays illuminate the sensor, they may be absorbed in the depletion region, allowing all photoelectrons to be collected by one pixel. This is why we observed a significant number of single-pixel events in Sec.~\ref{sect:xray_results} despite the significant level of observed crosstalk.

In astronomical imaging, the PSF delivered at the sensor will be convolved with the sensor's inherent PSF from crosstalk. The PSF from crosstalk, pictured for blue light in Fig.~\ref{fig:coupling_heatmaps}, was well fit by a Gaussian with FWHM around $1.4$--$1.6\,pix$, depending on the wavelength.

\begin{table}[t]
\centering
\begin{tabular}{c|c|c|c|c}
COSMOS Mode & Exposure Time (s) & Blue Light & Red Light & NIR Light \\
& & ($400\,nm$) & ($625\,nm$) & ($875\,nm$) \\
\hline
 & 0.1 & $48.4\pm 2.2\%$ & $56.4\pm 6.5\%$ & $43.8\pm 1.8\%$\\ 
CMS & 0.5 & $50.3\pm 0.5\%$ & $56.5\pm 1.0\%$ & $41.9\pm 1.2\%$ \\ 
 & 1 & $50.1\pm 0.8\%$ & $55.0\pm 0.4\%$ & $41.8\pm 0.4\%$ \\ \hline
& 0.1 & $46.1\pm 5.1\%$ & $62.5\pm 6.5\%$ & $46.1\pm 8.4\%$ \\ 
HSHGRS & 0.5 & $48.0\pm 1.9\%$ & $56.7\pm 2.6\%$ & $40.4\pm 0.5\%$ \\ 
 & 1 & $48.7\pm 0.6\%$ & $54.7\pm 0.1\%$ & $40.3\pm 1.0\%$ \\
\end{tabular}
\vspace{0.01\textwidth}
\caption{The fraction of signal recorded in the central (illuminated) pixel relative to the total signal recorded in an $11\times 11$ region of interest, for different wavelengths of light used, exposure times, and operating modes. Standard deviations across three images provided measurement uncertainties.}
\label{tab:crosstalk_data}
\end{table}

To detect IPC-induced crosstalk, we followed the methods of Refs.~\citenum{moore:2006} and \citenum{bottom:2024} of measuring pixel-pixel autocorrelations. We collected 24 uniformly illuminated images using the apparatus described in Sec.~\ref{sect:flats_apparatus}. From these, we generated 12 difference images and computed a $5 \times 5$ autocorrelation matrix. With this number of difference images and the resolution of the sensor, we could expect to detect IPC at levels as low as 100 parts per million. The fractional capacitive coupling at a given separation was determined by halving the ratio of the autocorrelation at that separation to the autocorrelation of a pixel with itself. Fig.~\ref{fig:coupling_heatmaps}b shows the measured magnitude of capacitive coupling. The presence of non-negligible capacitive coupling introduced a systematic underestimation of the conversion gain by approximately $1.6\%$ when using the mean-variance method. This factor has been accounted for in the conversion gain values presented in Table~\ref{tab:gain}. However, the coupling to nearest-neighbor pixels was measured to be below $0.3\%$, indicating that nearly all of the observed crosstalk likely arose from diffusion.

\subsection{Image Lag}
\label{sec:image_lag_results}
If the pixels in a CMOS image sensor do not fully transfer their collected charge for readout after an exposure, some electrons may appear in subsequent frames as ghosts of bright features, in a process referred to as image lag \cite{james:2021,bonjour:2012}. Image lag can undermine the photometric precision of a camera, especially if bright and dark objects are captured in sequence on the same pixel, such as during dithering or nodding for sky subtraction in ground-based astronomy. To determine whether the COSMOS sensor undergoes image lag, we compared sequences of dark frames before and after exposing the sensor to light. We did this by capturing 5 frames with the shutter closed, 5 frames with the shutter open, and 5 frames with the shutter closed once more, at a range of exposure times. The illumination flux was such that a 10 second exposure just saturated the pixels.

\begin{figure}[h!]
    \centering
    \includegraphics[width=0.7\linewidth]{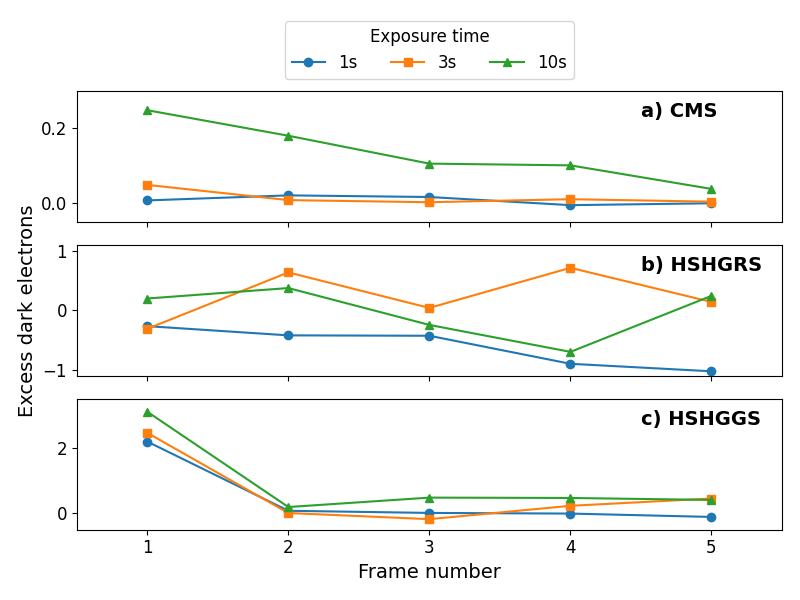}
    \caption{Excess number of electrons measured in five dark frames after capturing five bright frames for each mode.}
    \label{fig:Image_Lag}
\end{figure}

Figure~\ref{fig:Image_Lag} shows the excess brightness in dark frames after light exposure for each mode. In the CMS mode, an image lag of approximately 1/4 electron is seen with 10 second exposures that decays in approximately one minute. In the high-speed modes the image lag is always below the read noise except for the first frame in global shutter mode that shows 2-3 electrons excess, with weak dependence on the exposure time. If the sensor's nonlinearity was caused by charge being trapped in each pixel, then this trapped charge is released and cleared during pixel reset. While measurable, the amount of image lag shown by COSMOS is so low that it is unlikely to affect astronomical observations.

\section{On-Sky Demonstration}
\begin{figure}[h!]
    \centering
    \includegraphics[width=0.7\linewidth]{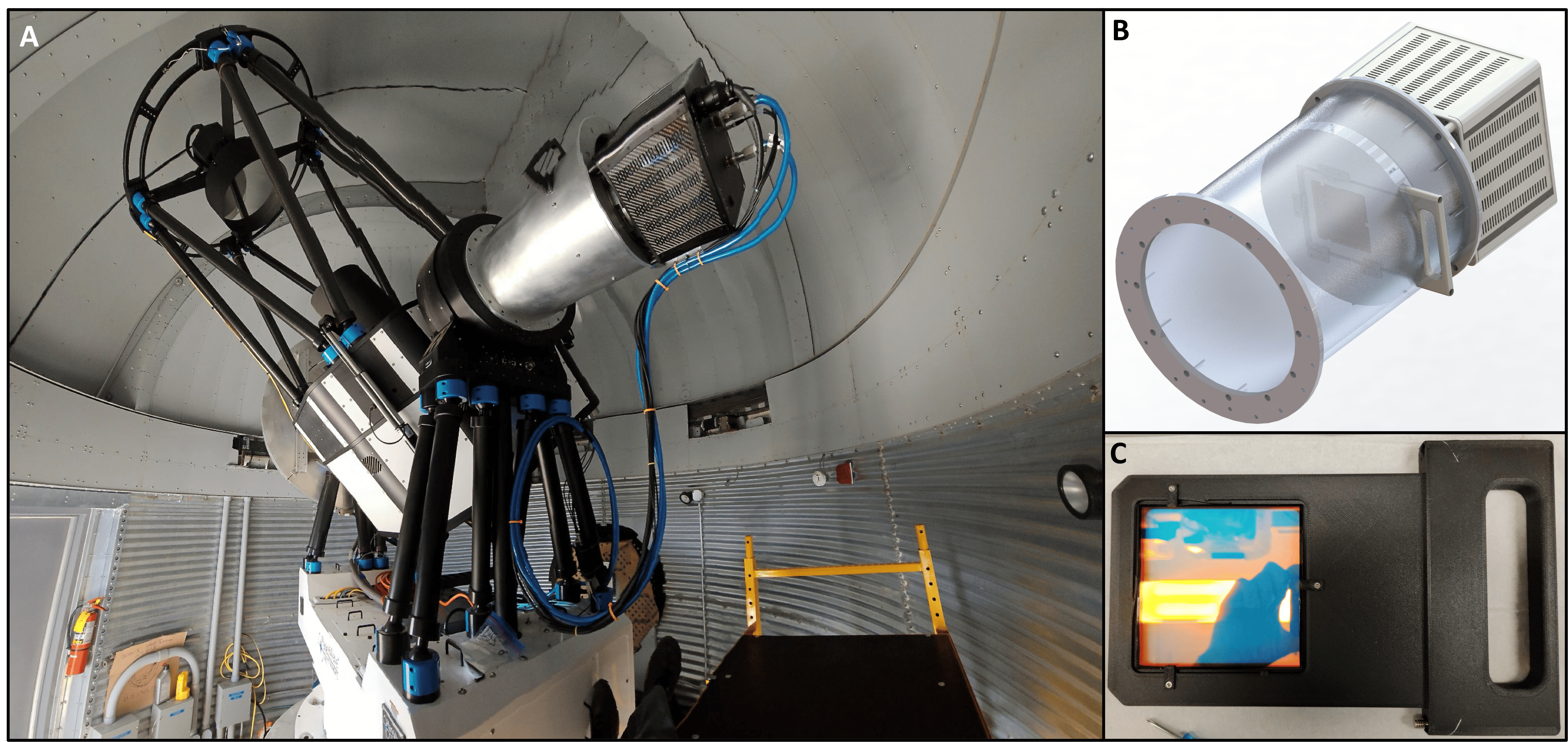}
    \caption{A) COSMOS mounted on the $1\,m$ WINTER telescope \cite{lourie:2020} at Palomar Observatory. B) A transparent CAD rendering of the simple telescope adapter, revealing the filter slot and the location of the 3D-printed tray. C) Close-up photo of a filter secured in one of the 3D-printed trays, used to manually swap filters.}
    \label{fig:cosmos_winter}
\end{figure}

To demonstrate the capabilities of the COSMOS camera for astronomical imaging, we installed it on a $1\, m$ telescope at Palomar Observatory. This telescope, an F/6 robotic Planewave CDK-1000 system, permanently hosts the Wide-Field Infrared Transient Explorer (WINTER)\cite{lourie:2020,frostig:2024} instrument but has a second Nasmyth port available for visitor instruments. In Fig.~\ref{fig:cosmos_winter}, the COSMOS camera is shown mounted on this Nasmyth port with a custom engineered mount and manual filter exchanger, which provides repeatability due to its preloaded kinematic support. To minimize internal reflections, all interior surfaces of the mount were lined with black flocked adhesive paper (Thorlabs BFP1). We equipped the system with three filters---Sloan $u'$, $g'$, and $r'$ housed in dedicated trays---sourced from Baader Planetarium in a $100\times100~mm^2$ format to avoid vignetting the COSMOS sensor. With this setup, the COSMOS camera provided a field of view of approximately $45\times 45\,arcmin^2$, with a pixel scale of $0.32\,arcsec/pixel$.

\begin{figure}[h!]
    \centering
    \includegraphics[width=0.7\linewidth]{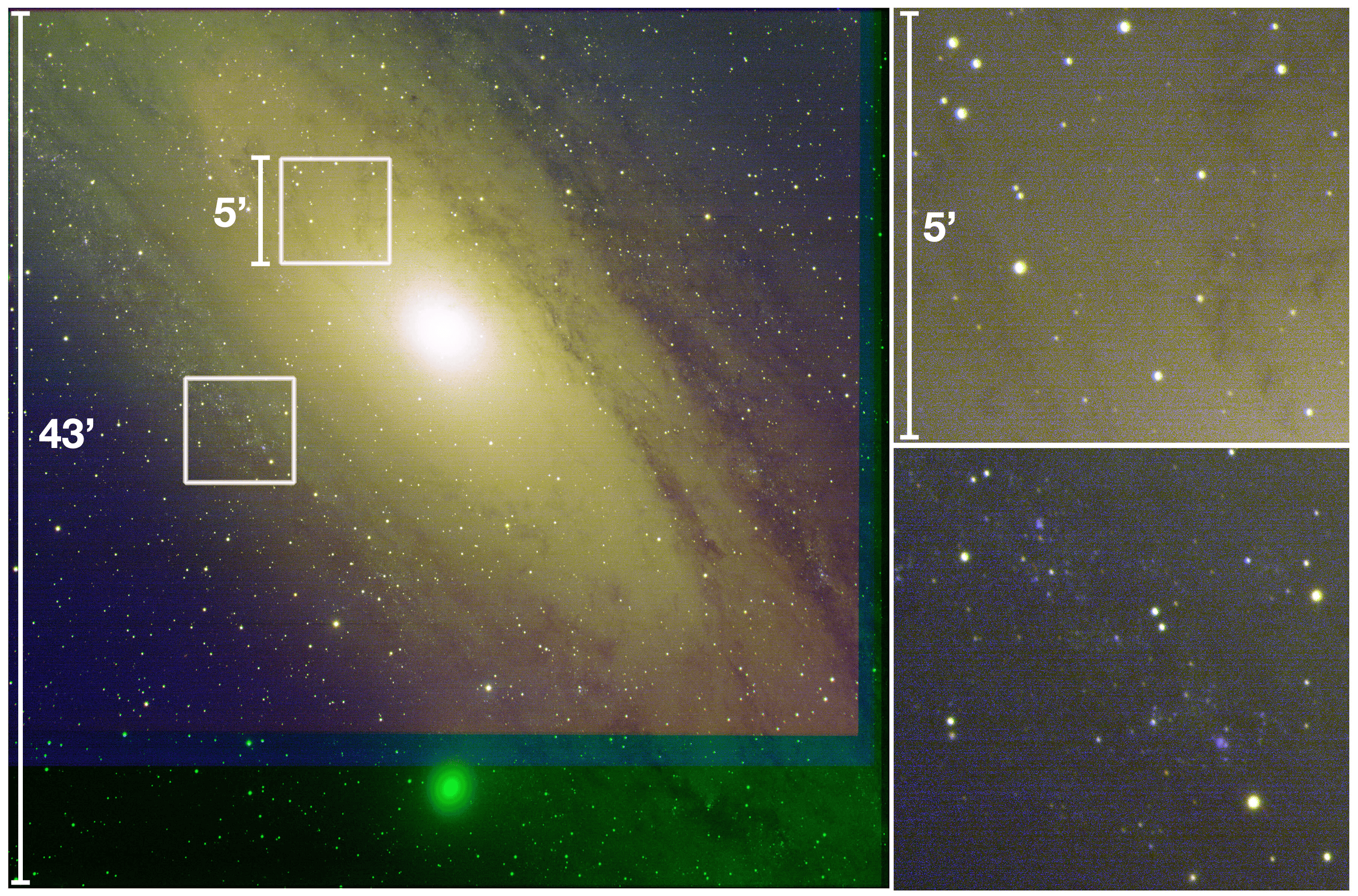}
    \caption{Stacked image of the Andromeda Galaxy taken with COSMOS on the WINTER telescope, using Sloan $u'$, $g'$, and $r'$ filters. Errors in telescope pointing (between the manual filter changes) led to misalignment in images taken with each filter. Zoomed-in boxes show $1000\times 1000$ pixel regions.}
    \label{fig:andromeda_fig}
\end{figure}

We first imaged the Andromeda Galaxy with each of the three filters, with an exposure time of $60\,s$ and a total duration of $10\,min$ for each filter. Figure~\ref{fig:andromeda_fig} shows the resulting stacked image, demonstrating the large field of view and high resolution deliverable by COSMOS. Images taken with each filter were offset due to pointing offsets as for manual filter exchanges we had to reposition the telescope.

\begin{figure}[h!]
    \centering
    \includegraphics[width=0.98\linewidth]{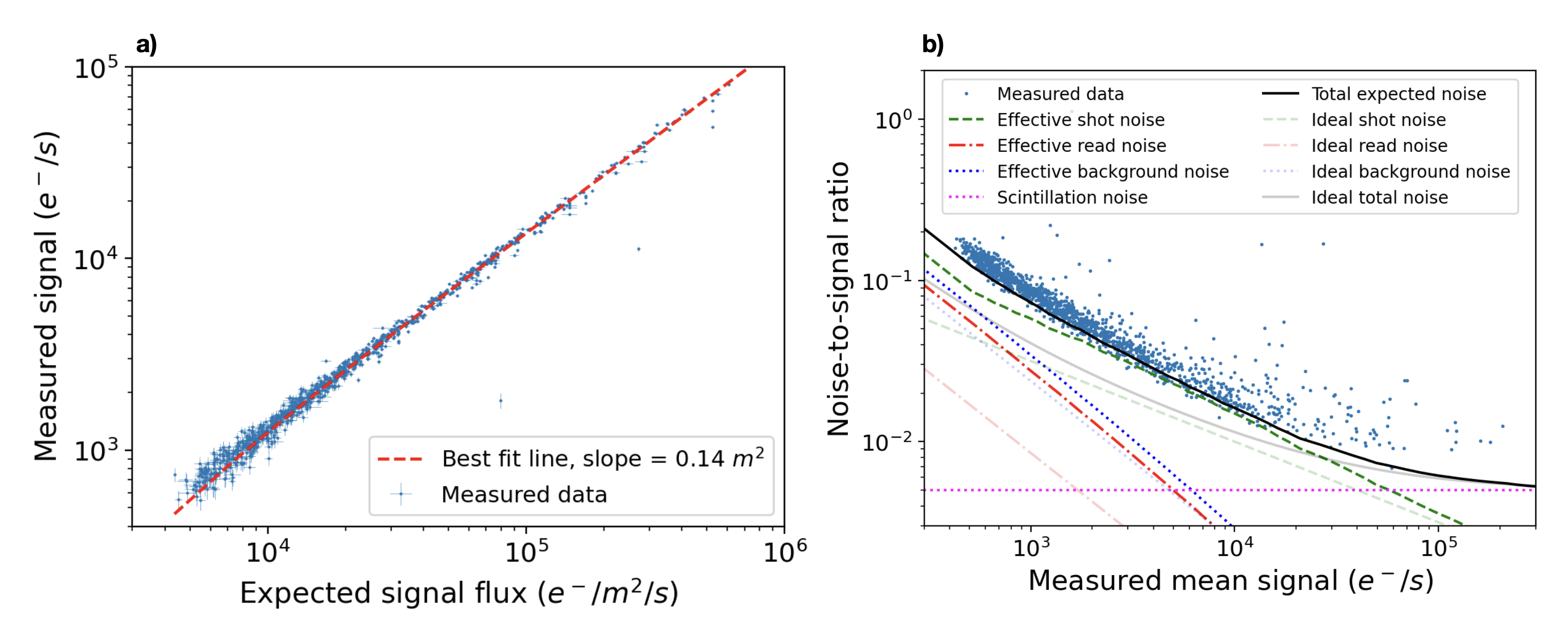}
    \caption{\textbf{a)} The measured signal in a COSMOS image for stars identified in an image of M38 vs.\ the expected signal flux of these stars, with best fit line in red dashes. \textbf{b)} Noise measured in light curves of identified stars vs.\ star brightness. The black solid line shows the total expected noise level, comprising effective shot noise (dark green dashed line), effective read noise (dotted blue line), effective background noise (dashed-dotted red line), and scintillation noise (dotted pink line). Noise components expected for an ideal sensor with no nonlinearity and a read noise of $0.7\,e^-$ are shown in semi-transparent curves.}
    \label{fig:on_sky_data}
\end{figure}

To test whether COSMOS with our nonlinearity calibration could deliver accurate and precise photometry, we observed the open cluster Messier 38 (M38) through the $r'$ filter, using the CMS mode, with an exposure time of $1\,s$, for a duration of 3 minutes. We cross-matched stars in this image with stars that have spectra recorded in the Gaia DR3 catalog \cite{gaia:2016,gaiaDR3:2023}. By multiplying the photon fluxes in these spectra with the bandpass of the filter and the measured COSMOS quantum efficiency, we calculated the total number of electrons expected for each star per aperture area. Hence, we did not need to perform additional observations for measuring extinction and color coefficients for a conventional standard transformation. We performed aperture photometry with one of our observed frames and compared the measured signal to the expected signal for each star, as shown in Fig.~\ref{fig:on_sky_data}a. We observed linear response with a slope of $0.14\,m^2$ corresponding to the effective area of the telescope. This value matched what has been observed with other cameras on the telescope and is the result of central obscuration, dust accumulation on the mirrors, and losses through the optical train. Almost every star fell along this line, to within measured errors. The horizontal error bars indicate flux errors reported by Gaia. The vertical error bars indicate the expected total noise level, including effective read noise, effective shot noise, effective background noise, and scintillation noise (calculated to be $0.5\%$\cite{young:1967}). We estimated the effective shot noise, read noise, and background noise by calculating the level of nonlinearity correction applied to every pixel in a star's PSF and scaling the noise values appropriately. This yielded an empirical relationship between a star's measured signal and its expected noise levels, which is valid as long as the PSF remains stable.

\label{sec:on_sky_results}

We also constructed light curves for all stars identified in the image (not just those bright enough to have Gaia spectra). As the brightnesses of these stars do not vary appreciably on the timescale of these light curves (a few minutes, during which color terms can also be ignored), and atmospheric variations were corrected using the mean brightnesses of all monitored stars, we measured the noise-to-signal ratio for each star by taking the standard deviation and mean value of its light curve. We show the measured noise-to-signal ratio vs.\ mean brightness in Fig.~\ref{fig:on_sky_data}b. These data agreed well with the expected noise level, plotted in the black solid line. The effective shot noise, effective read noise, and effective background noise curves were calculated using the empirical relationship described above. In semi-transparent curves, we show the shot noise, read noise, background noise, and total noise expected if no nonlinearity calibration would be required. For most stars, this ideal total noise is approximately half of the actual total noise, highlighting the impact the sensor's nonlinearity has on photometric precision.

Finally, to study the effect of crosstalk on image quality, we compared the observed PSFs to the seeing measured at Palomar Observatory. We measured the FWHM of the PSFs of the brightest 100 stars in one image of M38 to be $1.6\pm 0.2\,arcsec$, or approximately $5$ pixels. This agreed with the FWHM of the atmospheric seeing measured at the time of our observations at the 200-inch Hale telescope at Palomar, $1.4$--$1.7\,arcsec$. This demonstrated that for oversampled imaging, the pixel crosstalk has little effect on the PSF, as expected for a sensor PSF with a FWHM of approximately $1.5$ pixels.

\section{Conclusion}
\label{sect:conclusion}
We conducted benchtop tests and an on-sky demonstration with the Teledyne COSMOS-66 camera. The raw response of the camera exhibited significant nonlinearity in the low-light regime, plausibly caused by the trapping of electrons in each pixel. We developed a method for calibrating the responses of all pixels in the COSMOS image sensor, which produced linear response across the dynamic range of the sensor. This pixel-by-pixel calibration also removed nearly all fixed-pattern noise, yielding a pixel response nonuniformity (PRNU) value of $0.09\%$. As a result of the nonlinearity and the calibration to address it, the sensor's read noise and shot noise at small signal levels are scaled up. The effective read noise of the sensor for the lowest signal levels was $2.9\,e^-$ RMS for the CMS mode, though this value is expected to taper to $0.77\,e^-$ RMS when the raw sensor response becomes linear (above $1000\,e^-$). Similarly, the effective shot noise for a small number of collected electrons $N_e$ is approximately $2\sqrt{N_e}$, tapering to the expected $\sqrt{N_e}$ in the linear regime.

The dark current in the sensor at $-25^\circ C$ was $0.12\,e^-/pix/s$. The quantum efficiency of the sensor exceeded 50\% for $250\,nm$--$800\,nm$ light and reached above 85\% at $600\,nm$. The sensor showed a significant amount of crosstalk due to electron diffusion. Due to this crosstalk, the inherent PSF of the sensor could be approximated as a Gaussian with FWHM of $\approx 1.5\,pix$. For under- or critically sampled images, the crosstalk may therefore noticeably broaden the PSF. We demonstrated the performance of the COSMOS camera on the $1\,m$ WINTER telescope \cite{lourie:2020} at Palomar Observatory and verified that the sensor, when calibrated, delivers accurate photometric measurements. The noise observed in our astronomical imaging was consistent with that expected from the effective read noise and shot noise.

With its large pixels, high resolution, and high frame rate, the COSMOS camera is architecturally suited to large scale, time domain astronomical surveys. However, because its nonlinear response effectively scales up noise at low signal levels, COSMOS does not currently provide a clear advantage over large-format CCDs or mosaics of low-noise CMOS image sensors. Improvements to the sensor's pixel design are already underway at Teledyne \cite{Teledyne2025}. If these successfully eliminate the sensor's nonlinearity and crosstalk, then the camera could achieve sub-electron read noise and deliver exceptional photometric precision across a wide dynamic range. In this case, the camera would combine the advantages of large-format CCDs and CMOS image sensors and could be a workhouse instrument for many astronomical science cases.

\appendix    

\subsection*{Disclosures}
The authors acknowledge technical support provided by Teledyne Digital Imaging throughout this project. Teledyne also shared information regarding future development plans related to their camera systems. Additionally, aspects of the results presented in this work were discussed with Teledyne prior to publication. However, Teledyne did not have any role in the final analysis, the interpretation of the data, or the conclusions drawn in this paper. While our team has been given, under a non-disclosure agreement, access to proprietary data that allowed us to understand the device in depth, this paper does not contain any confidential material, intellectual property, or controlled information. We declare no other financial, commercial, or other conflicts of interest related to the research presented in this paper.

\subsection*{Code, Data, and Materials Availability}
The data generated and analyzed in this study are not publicly available due to their large volume. However, the code and a representative subset of the data used for this work are available from the corresponding author upon reasonable request.

\subsection* {Acknowledgments}
We thank Teledyne Digital Imaging for loaning us a demonstration unit COSMOS camera for laboratory characterization and on-sky demonstration, and also for providing extensive technical support. We particularly appreciate how forthcoming and open Teledyne has been with technical discussions, providing us with direct access to their engineering teams on short timelines to track down and resolve issues. We acknowledge Mic Chaudoir, Craig Wall, and Sebastian Remi from the Teledyne team for their invaluable support in the receipt and operation of the camera. The work was funded in part by the MIT Kavli Research Investment Fund (grant MKI KRIF 2654353). Chris Layden warmly thanks George Ricker for his support and advisement throughout this project. We thank Kevin Burdge for his advice regarding on-sky observation and reduction of on-sky data. We thank Erin Kado-Fong, observer on the Palomar Observatory P200 telescope on the night of our on-sky demonstration, for providing measured values of the atmospheric seeing.

This research made use of Photutils, an Astropy package for
detection and photometry of astronomical sources (Bradley et al.
2024). This work has made use of data from the European Space Agency (ESA) mission
\textit{Gaia} (\url{https://www.cosmos.esa.int/gaia}), processed by the \textit{Gaia}
Data Processing and Analysis Consortium (DPAC,
\url{https://www.cosmos.esa.int/web/gaia/dpac/consortium}). Funding for the DPAC
has been provided by national institutions, in particular the institutions
participating in the \textit{Gaia} Multilateral Agreement. 


\bibliography{references.bib}   
\bibliographystyle{spiejour}   


\vspace{2ex}\noindent\textbf{}

\vspace{1ex}
\noindent

\listoffigures
\listoftables

\end{spacing}
\end{document}